\documentclass[reprint, superscriptaddress, secnumarabic, amssymb, nobibnotes, aps, prl]{revtex4-1}

\usepackage{graphicx}
\usepackage{epstopdf}
\usepackage[T1]{fontenc}
\usepackage[latin9]{inputenc}
\usepackage{amsbsy}
\usepackage{gensymb}
\usepackage[T1]{fontenc}
\usepackage[latin9]{inputenc}
\usepackage{amsmath}
\usepackage{amssymb}
\usepackage{bbm}
\usepackage{physics}
\usepackage{braket}
\usepackage{xcolor}
\allowdisplaybreaks
\usepackage{graphicx}
\usepackage[colorlinks=true]{hyperref}  
\hypersetup{
    bookmarks=true,         
    unicode=false,          
    pdftoolbar=true,        
    pdfmenubar=true,        
    pdffitwindow=false,     
    pdfstartview={FitH},    
    pdftitle={Nb1.7Ta3.3S2},    
    pdfauthor={},     
    pdfsubject={},   
    pdfcreator={},   
    pdfproducer={}, 
    pdfkeywords={} {} {}, 
    pdfnewwindow=true,      
    colorlinks=true,       
    linkcolor=blue, 
    citecolor=blue,        
    filecolor=magenta,      
    urlcolor=blue           
} 
\usepackage[normalem]{ulem}

\newcommand{\equref}[1]{Eq.~(\ref{#1})}

\newcommand{\figref}[1]{Fig.~\ref{#1}}

\newcommand{\tableref}[1]{Table~\ref{#1}}

\begin{document}
\title{Superconducting properties of van der Waals metal-rich subsulfide Nb$_{1.7}$Ta$_{3.3}$S$_2$}
\author{A. Kataria}
\affiliation{Department of Physics, Indian Institute of Science Education and Research Bhopal, Bhopal, 462066, India}
\author{R.~P.~Singh}
\email[]{rpsingh@iiserb.ac.in} 
\affiliation{Department of Physics, Indian Institute of Science Education and Research Bhopal, Bhopal, 462066, India}

\begin{abstract}

Layered metal-rich subsulfides have become a promising area for exploring intriguing properties such as superconductivity, nontrivial topology, and charge density waves; however, despite their extraordinary potential, they have remained largely unexplored. We report a comprehensive analysis of a van der Waals layered metal-rich subsulfide, Nb$_{1.7}$Ta$_{3.3}$S$_2$, using a range of measuring techniques, including AC transport, magnetization, and specific heat. Our measurements confirm the occurrence of a type-II superconducting transition at a temperature of approximately $T_c$ = 3.64(1) K. Furthermore, specific heat measurements suggest weakly coupled, nodeless superconductivity.

\end{abstract}

\maketitle
\section{Introduction}

The structural and chemical diversity of transition metal chalcogenides (TMXs) provides a vast area of study in condensed matter physics and exciting research opportunities \cite{tmc}. Although metal dichalcogenides (TMDs) have been extensively studied for their nontrivial band topology and exotic properties, metal-rich chalcogenides have yet to be explored, presenting an open field for investigating their properties and characteristics. Metal-rich chalcogenides exhibit a wide range of crystal structures, rich crystal chemistry, unusual electronic properties, and superconductivity, indicating their potential for multiple applications \cite{mr,mr2,mr3,mr4,n21s8}.

In particular, layered metal-rich chalcogenides are an attractive system for studying due to their layered crystal structure, which can contribute to anisotropic, layered-dependent properties and nontrivial topological states \cite{ans,lyp,b2s3}. Superconducting layered systems, such as cuprates, Fe-based, or Bi-based superconductors, are also known for their unconventional behaviour, and high transition temperature \cite{cu,fe,bi,mb2}. Further, in metal-rich TMXs, theoretical calculations show that metal-nonmetal bonding strongly impacts structural variation and electronic band structure \cite{bandstr}. The presence of high Z metals in metal-rich TMXs can enhance spin-orbit coupling (SOC) significantly, influencing the associated quantum ground state and can lead to the unconventional nature and nontrivial topology of the electronic band structure \cite{soc,soc1,soc2}. Therefore, studying the superconductivity in a layered system with strong metal interaction could be exciting for exploring the superconducting ground state and their relation. It makes layered metal-rich TMXs an interesting reference system for research.

\begin{figure*} 
\includegraphics[width=2.0\columnwidth, origin=b]{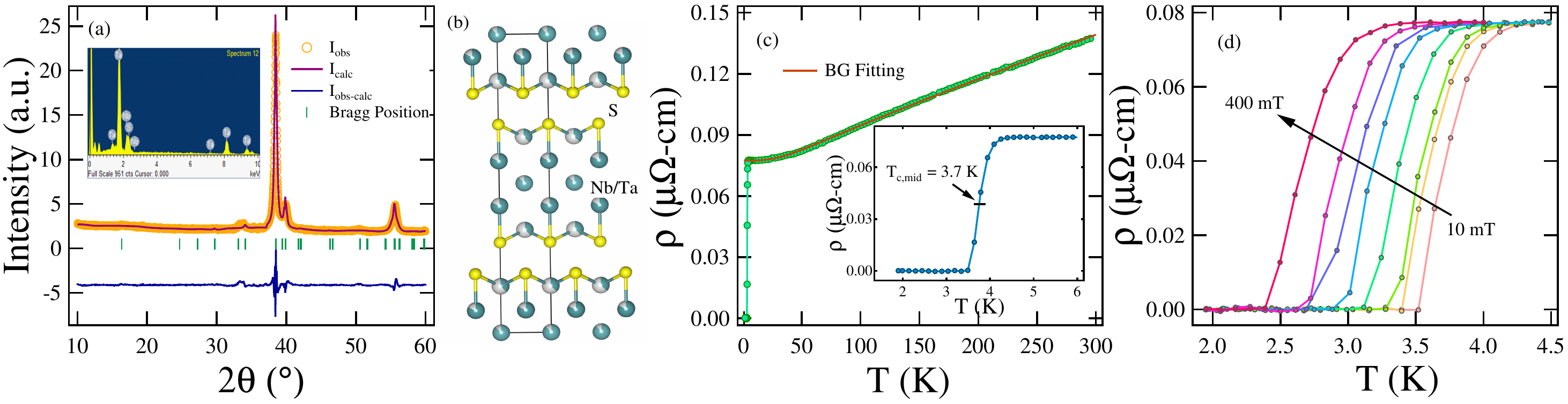}
\caption{\label{Fig1:Xrd} (a) Reitveld refined powder XRD pattern of polycrystalline Nb$_{1.7}$Ta$_{3.3}$S$_2$ with the inset showing the EDAX spectra. (b) Nb$_{1.7}$Ta$_{3.3}$S$_2$ crystal structure with yellow balls as sulfur atoms and green/grey balls presenting an equal probability of occupying Nb and Ta atoms. (c) $\rho(T)$ temperature-dependence with inset representing a sharp resistivity drop from the normal state. (d)  $\rho(T)$ versus $T$ curve at different applied magnetic field. }
\end{figure*}

Subsulfides, a class of metal-rich chalcogenide, have recently emerged as a new playground for exploring exotic properties such as nontrivial topological phases, charge density waves, and superconductivity \cite{tsm,wsm, brs, brs2, t2se,r9i4s4,bns}. They are named subsulfides due to their lower sulfur content than conventional, more ionic sulfides and exhibit properties that fall between intermetallic and ionic phases. However, despite their remarkable properties, these materials are largely unexplored due to their challenging synthesis, as they can form multiple phases. Nevertheless, their complex nature and potential for high SOC make them ideal for investigating novel superconductors. Studies on more superconducting subsulfides are important for a deeper understanding of the superconducting nature of these materials. In this regard, we study Nb$_x$Ta$_{5-x}$S$_2$ ($x$ = 1.7), a pseudobinary Ta$_5$Se$_2$ isostructural system. The structure consists of a metal-metal layer similar to the BCC structure, and a sulfur layer at the end connects another building block with the van der Waals interaction (\figref{Fig1:Xrd}(b)), which is completely different from niobium and tantalum sulfides TMDs structure \cite{nts,ntssc}. Septuple layers, S-M1-M2-M3-M2-M1-S (M = Nb/Ta), can also be viewed as MS$_2$ layered TMDs having S-M-S layers intercalated by additional metal layers. Intercalation with different layered materials (for example, 4$Hb$-TaS$_2$) or organic molecules in NbS$_2$ and TaS$_2$ TMDs has been well known to enhance superconductivity with the possibility of nontrivial phase \cite{4hbts2,otas2, ns2_2, ns2_3, ts2,ts2_3}. Due to its possible high SOC, intriguing intercalated layered structure and van der Waals interaction with metal-metal and metal-nonmetal bonding, it has the potential to exhibit unique properties, motivating a detailed study of this system. Furthermore, chalcogenides rich in metal and containing superconducting metal layers may serve as a prototype for creating novel superconducting materials featuring metallic areas separated by van der Waals layers. This arrangement could facilitate the exploration of the superconducting diode effect, proposed in systems comprising alternating layers of superconducting and non-superconducting materials or low-dimensional superconducting layered materials possessing high SOC \cite{sde,sde2,sde3,sde4,sde5}.

In this paper, the superconductivity in van der Waals metal-rich subsulfide Nb$_x$Ta$_{5-x}$S$_2$ ($x$ = 1.7) [where metal bonded Nb-Ta sheet is separated by sulfur layer] is detailed examined using x-ray diffraction, AC transport, magnetization, and specific heat measurements. All the corresponding measurements suggest bulk type-II superconductivity with $T_c$ = 3.64(1) K. The metallic nature in normal-state and weakly coupled superconducting state are also characterised, along with their respective electronic and superconducting parameters. 
 
\section{Experimental Details}

The polycrystalline sample of Nb$_{1.7}$Ta$_{3.3}$S$_2$ was prepared by arc melting the high purity constituent elements Nb(99.9\%), Ta(99.99\%) and S(99.9999\%) under Argon atmosphere. A 20\% extra S from the stoichiometric ratio was taken to maintain its loss during the melting process. The ingot was flipped several times while melting to ensure phase homogeneity. A powder x-ray diffraction (XRD) pattern was recorded for an as-cast sample at room temperature using a PANalytical diffractometer equipped with Cu K$_{\alpha}$ radiation ($\lambda$ = 1.54056 \AA). An oxford instrument's scanning electron microscope (SEM) is used for energy-dispersive x-ray analysis (EDAX). Magnetic measurements were performed on a superconducting quantum interference device (MPMS, Quantum Design). AC transport measurements using the four-probe technique and specific heat measurements using the two-$\tau$ relaxation method were performed on a 9 T physical property measurement system (PPMS, Quantum Design). 

\section{Results and Discussion}

\subsection{Sample characterization}

The powered XRD pattern was Rietveld refined using FullProf software, as shown in \figref{Fig1:Xrd}(a) \cite{fp}. Refinement confirms crystallization in the tetragonal structure with space group $I4/mmm$ and lattice constant $a$ = $b$ = 3.3089(6) \text{\AA} and $c$ = 21.6584(3) \text{\AA}, which is in agreement with previous reports \cite{nts}. Nb$_{1.7}$Ta$_{3.3}$S$_2$ structure contains five Nb-Ta (BCC type) layers and metal-chalcogenides layer blocks, as shown in the \figref{Fig1:Xrd}(b). The elemental analysis of the polycrystalline system from the EDAX spectra (inset of \figref{Fig1:Xrd}(a) indicates the nominal composition and the presence of sulfur within the experimental error limit.

\subsection{Superconducting and normal state properties}

\subsubsection{Electrical Resistivity}
\begin{figure*} 
\includegraphics[width = 2\columnwidth, origin=b]{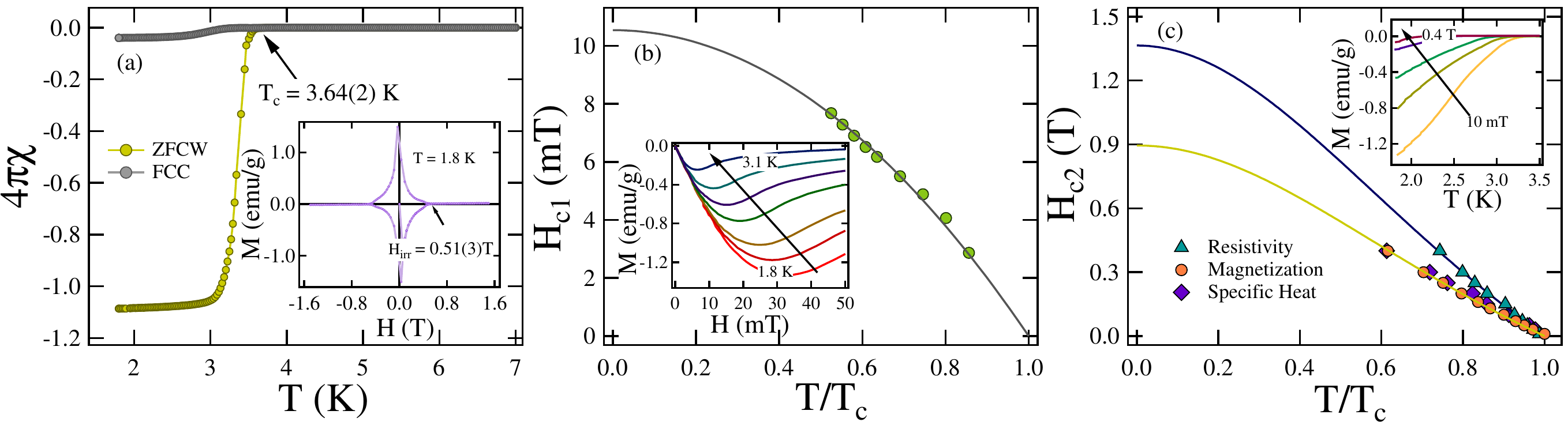}
\caption{\label{Fig2:Mag} (a) Temperature-dependent magnetic susceptibility showing $T_c$ = 3.64(2) K and inset presenting the magnetization loop. (b) Temperature dependence of $H_{c1}$ for Nb$_{1.7}$Ta$_{3.3}$S$_2$. The inset represents the low-field magnetization curves at various temperatures. (c) The upper critical field temperature variation estimated from specific heat, resistivity, and magnetization data, where solid lines represent the fitting using \equref{eqn3:Hc2} and inset displays the $M$ versus $T$ at different applied fields.}
\end{figure*}
The temperature-dependent AC resistivity of Nb$_{1.7}$Ta$_{3.3}$S$_2$ under zero magnetic field is shown in \figref{Fig1:Xrd}(c). The linear temperature dependence of resistivity up to high temperature depicts the metallic nature of the sample. A sharp drop in resistivity indicating superconductivity is observed, and its 50\% drop to the normal state value is  at $T_{c,mid}$ = 3.7(2) K with a transition width of 0.7 K. The measured superconducting transition temperature matches well with the previous study \cite{ntssc} and is distinct from the transition temperature (5.3 K) of alloy Nb$_{1.7}$Ta$_{3.3}$ (more details in appendix) \cite{nbta,nbta1}. Whereas the superconducting transition of single crystalline TaS$_2$ and NbS$_2$ is 0.8 K and 6.2 K, respectively \cite{tas2_sc,nbs2_sc}. The normal-state resistivity data in the above $T_c$ temperature range can be well understood by the Bloch-Gr\"uneisen (BG) model, which accounts for the scattering of electrons from the acoustic phonons \cite{bg}. To analyze the metallic phase of Nb$_{1.7}$Ta$_{3.3}$S$_2$, \equref{eqn1:rho} is used to fit the $\rho (T)$. 
\begin{equation}
\rho (T) = \rho_{0} + C\left(\frac{T}{\theta_R}\right)^3 \int_{0}^{\theta_R/T}\frac{x^3}{(e^x-1)(1-e^{-x})}dx
\label{eqn1:rho}
\end{equation} 
where the second term describes the BG model and $\rho_0$ is the residual resistivity due to defect scattering or disorder. $\theta_{R}$ is the Debye temperature, and $C$ is the material-specific parameter \cite{c}. A fit to the data, shown in \figref{Fig1:Xrd}(c), provides $C$ = 2.39(4) $\mu\Omega$-cm, $\theta_{R}$ = 227(2) K and residual resistivity, $\rho_{0}$ = 7.73(1) $\mu\Omega$-cm. Furthermore, the residual resistivity ratio, RRR, is $\rho(295 K)/\rho(10 K)$ = 1.77, whose low value suggests the dominance of disorder and defects in the sample due to its polycrystalline nature. \figref{Fig1:Xrd}(d) presents the change in the superconducting transition temperature in resistivity measurement with increasing applied magnetic field.

\subsubsection{Magnetization}

\figref{Fig2:Mag}(a) shows the temperature-dependent magnetization curves from two different modes, zero-field-cooled warming (ZFCW) and field-cooled cooling (FCC), under an applied magnetic field of 1 mT. A strong diamagnetic signal in the ZFCW curve at the temperature, $T_c$ = 3.64(1) K, is observed. The weak magnetic signal in the FCC curve suggests significant flux pinning in Nb$_{1.7}$Ta$_{3.3}$S$_2$. Further, the superconducting fraction greater than 100\% is due to the irregular shape of the sample and demagnetization effects \cite{dm}. The magnetization loop at 1.8 K is shown in the inset of \figref{Fig2:Mag}(a), indicating a type-II superconducting nature with a small area under the curve. An irreversible magnetic field, $H_{irr}$, is noted at 0.51(3) T, defined as a field above which vortices start to de-pins.

The magnetization curve in the low magnetic field region measured at different temperatures up to the transition temperature determines the lower critical field, $H_{c1}$(0). The point of deviation from the linear fit in the magnetization curves (inset of \figref{Fig2:Mag}(b)) is defined as $H_{c1}$ for the respective temperature. With increasing temperature, the value of $H_{c1}$ decreases, and the relation is fitted using the Ginzburg-Landau (GL) equation,
\begin{equation}
H_{c1}(T) = H_{c1}(0) \left[1-\left(\frac{T}{T_{c}}\right)^{2}\right].
\label{eqn2:Hc1}
\end{equation}
This provides the lower critical field value, $H_{c1}(0)$ = 10.53(8) mT, as shown in \figref{Fig2:Mag}(b). The upper critical field, $H_{c2}(0)$, is extracted from the temperature-dependent measurements at different applied magnetic fields. \figref{Fig1:Xrd}(d) and inset of \figref{Fig2:Mag}(c) depict the effect of increasing applied magnetic field on the transition temperature in resistivity and magnetization measurements, respectively. The obtained $H_{c2}$ values as a function of reduced temperature $t = T/T_c$ is plotted in \figref{Fig2:Mag}(c), and its temperature dependence is analyzed by combining the semi-empirical and GL relation,
\begin{equation}
H_{C2}(T) = H_{C2}(0)\left[\frac{1-t^2}{1+t^2}\right]. 
\label{eqn3:Hc2}
\end{equation}
The fitting from different measurements, including resistivity, magnetization and specific heat, yield $H_{c2}(0)$ = 1.36(1) T, 0.89(1) T, 0.92(2) T, respectively. 

From BCS theory, the upper critical field of a superconductor is limited by either the orbital critical field or Pauli limiting field. The Werthamer-Helfand-Hohenberg (WHH) theory of a type-II superconductor can be used to estimate the orbital limit of $H_{c2}$ from the expression \cite{WHH_1, WHH_2},
\begin{equation}
 H_{c2}^{Orb}(0) = -\alpha T_{c}\left.\frac{dH_{c2}(T)}{dT}\right|_{T=T_{c}}
\label{eqn3:HHH}
\end{equation}
where $\alpha$ is the purity factor considered 0.69 for the dirty superconductor and 0.73 for the clean limit superconductor. For Nb$_{1.7}$Ta$_{3.3}$S$_2$, the initial slope $\frac{-dH_{c2}(T)}{dT} $ at $T = T_{c}$ is estimated to be 0.25(1) T/K. For $\alpha$ = 0.69, the calculated orbital limiting field, $ H_{c2}^{Orb}(0)$ = 0.63(3) T. Furthermore, the Pauli limiting field is given by $ H_{c2}^{P}(0)$ = const.$T_{c} $, where const. = 1.86 T/K \cite{Pauli_1,Pauli_2}. For Nb$_{1.7}$Ta$_{3.3}$S$_2$, $ H_{c2}^{P}(0)$ = 6.77(2) T for $T_c$ = 3.64 K. The upper critical field values are significantly lower than the Pauli limiting field, indicating that the orbital effect is responsible for Cooper pair breaking in Nb$_{1.7}$Ta$_{3.3}$S$_2$. 
 
The $H_{c1}(0)$ and $H_{c2}(0)$ values (from magnetization data) are further used to estimate the length parameters GL coherence length, $\xi_{GL}(0)$ and penetration depth $\lambda_{GL}(0)$, by the following relations \cite{Coh_Leng,lamda},
\begin{align}
\label{eqn4:lp}
\begin{split}
H_{c2}(0) &= \frac{\Phi_{0}}{2\pi\xi_{GL}^{2}(0)},
\\
H_{c1}(0) &= \frac{\Phi_{0}}{4\pi\lambda_{GL}^2(0)}\left(\mathrm{ln}\frac{\lambda_{GL}(0)}{\xi_{GL}(0)}+0.12\right)
\end{split}
\end{align}
where $\Phi_0$ is the magnetic flux quantum. From the above relation, the calculated $\xi_{GL}(0)$ = 19.2(2) nm and $\lambda_{GL}(0)$ = 195(3) nm. The GL parameter, $\kappa_{GL}$ is simply extracted from $\kappa_{GL} = \frac{\lambda_{GL}(0)}{\xi_{GL}(0)}$ and the obtained value is 10.1(2). The high value of $\kappa_{GL}$ suggests that Nb$_{1.7}$Ta$_{3.3}$S$_2$ is a strong type-II superconductor. Furthermore, the thermodynamic critical field, $H_{c}$ is calculated using the equation $H_{c}^2\ln{\kappa_{GL}} = H_{c1}(0)H_{c2}(0)$ \cite{lamda}, providing $H_{c}$ = 63(2) mT.

\subsubsection{Specific heat}

\begin{figure} 
\includegraphics[width=0.92\columnwidth, origin=b]{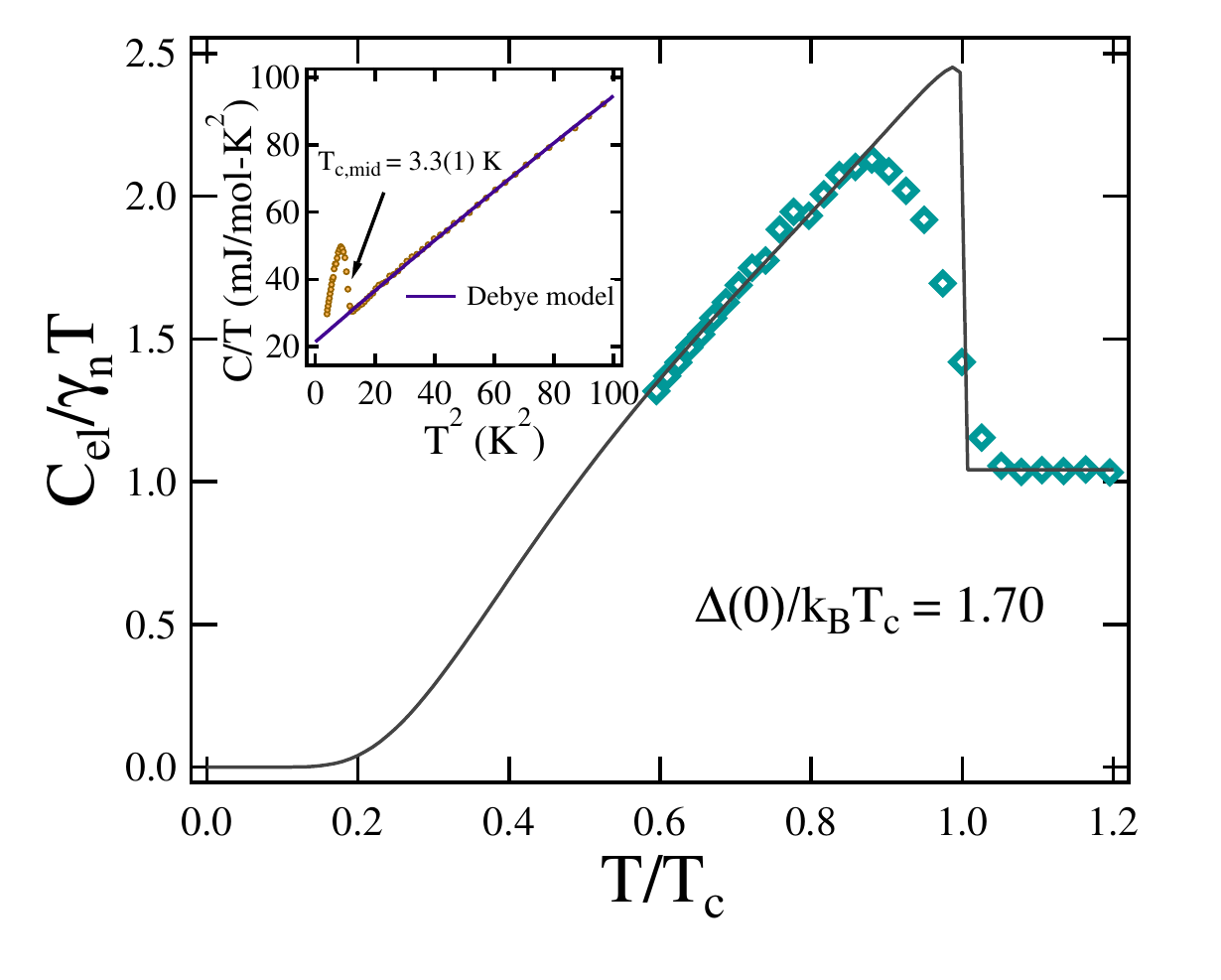}
\caption{\label{Fig3:SH} The normalized specific heat data, $C_{el}/\gamma_{n}T$, is well described by a single-gap s-wave model shown by the solid black line. Inset shows the $C/T$ versus $T^{2}$ data fit well with the Debye relation.}
\end{figure}

Bulk superconductivity in Nb$_{1.7}$Ta$_{3.3}$S$_2$ is further confirmed by measuring the specific heat at the zero field. In the inset of \figref{Fig3:SH}, a discontinuity (superconducting anomaly) is observed at the transition temperature, $T_{c,mid}$ = 3.3(1) K, which is in reasonable agreement with the other measurements. The normal-state specific heat data is fitted with the Debye model ${C/T}=\gamma_{n} +\beta_{3}T^{2}$, where the Sommerfeld coefficient $\gamma_{n}$ represents the electronic contribution and the Debye constant $\beta_{3}$ is the phononic part in the specific heat. The best fit of the data yield $\gamma_{n}$ = 21.7(2) mJmol$^{-1}$K$^{-2}$ and $\beta_{3}$ = 0.74(1) mJmol$^{-1}$K$^{-4}$ for Nb$_{1.7}$Ta$_{3.3}$S$_2$. 

Further, $\beta_{3}$ is used to evaluated the Debye temperature via relation \cite{thetad}, $\theta_{D} = \left(\frac{12\pi^{4}RN}{5\beta_{3}}\right)^{\frac{1}{3}}$, where $R$ = 8.314 J mol$ ^{-1} $K$ ^{-1} $ is a gas constant. The calculated value of $\theta_{D}$ = 263(3) K is close to the value obtained from resistivity measurement. The density of states at the Fermi level, $D_{C}(E_{\mathrm{F}})$, can be represented in terms of $\gamma_{n}$, as $\gamma_n  = \left( \frac{\pi^{2}k_{B}^{2}}{3}\right) D_{C}(E_{\mathrm{F}})$, where k$_{\mathrm{B}}$ = 1.38$\times$10$^{-23}$ J K$^{-1}$. $D_{C}(E_{\mathrm{F}})$ is estimated to be 9.23(8) states eV$^{-1}$f.u.$^{-1}$. Moreover, McMillan's theory determines the electron-phonon coupling constant, which estimates the strength of attractive interaction between electrons and phonons. The dimensionless coupling constant, $\lambda_{e-ph}$ is expressed as \cite{McMillan},
\begin{equation}
\lambda_{e-ph} = \frac{1.04+\mu^{*}\mathrm{ln}(\theta_{D}/1.45T_{c})}{(1-0.62\mu^{*})\mathrm{ln}(\theta_{D}/1.45T_{c}) - 1.04 }
\label{eqn5:Lambda}
\end{equation}
where $\mu^*$ is the repulsive screened Coulomb parameter and is typically equal to 0.13 for intermetallic compounds. From the estimated value of $ \theta_{D}$ = 263 K and measured value of $T_{c}$ = 3.3 K, $ \lambda_{e-ph} $ is calculated to be 0.59(2). The obtained value of $\lambda_{e-ph}$ classify Nb$_{1.7}$Ta$_{3.3}$S$_2$ as a weakly coupled superconductor.

The electronic contribution, $C_{el}$ of the specific heat, at zero applied field can be calculated using $C_{el}$  = $C$ - $\beta_{3}T^{3}$, to characterize the superconducting gap properties. \figref{Fig3:SH} shows the variation of normalised electronic specific heat, $C_{el}/{\gamma_{n}T}$, as a function of reduced temperature. The low-temperature normalized $C_{el}$ is analyzed using the $\alpha$-model of a single-gap BCS superconductor \cite{alp}. The model can be represented in terms of entropy, $S$, as,
\begin{equation}
\frac{S}{\gamma_{n}T_{C}} = -\frac{6}{\pi^2}\left(\frac{\Delta(0)}{k_{\mathrm{B}}T_{C}}\right)\int_{0}^{\infty}[ \textit{f}\ln(f)+(1-f)\ln(1-f)]dy 
\label{eqn6:BCS1}
\end{equation}
\\
where $f(\xi)$ = $[\exp(E(\xi)/k_{\mathrm{B}}T)+1]^{-1}$ is the Fermi function and $y$ = $\xi/\Delta(0)$. $E(\xi)$ is the energy of the normal electrons relative to Fermi energy, expressed as $E(\xi)$ = $\sqrt{\xi^{2}+\Delta^{2}(t)}$, where $t=T/T_c$ and $\Delta(t)$ = tanh[1.82(1.018(($\mathit{1/t}$)-1))$^{0.51}$] is the BCS approximation for the temperature dependence of energy gap. The electronic specific heat, $C_{el}$ is related to $S$ via $C_{el} = t dS/dt$. The best fit of the curve $C_{el}/\gamma_n T$ versus $T/T_c$ yields $\alpha$ = $\Delta(0)/k_{\mathrm{B}}T_{c}$ = 1.70(5), which is close to the BCS value in the weak coupling limit 1.76, suggesting weakly coupled superconductivity in Nb$_{1.7}$Ta$_{3.3}$S$_2$. The solid black line in \figref{Fig3:SH} represents the $C_{el}/\gamma_n T$ fitting with a single-gap s-wave (nodeless) BCS model.

\subsubsection{Electronic properties and the Uemura plot}

The electronic properties of the metal-rich subslfide Nb$_{1.7}$Ta$_{3.3}$S$_2$ have been quantified using the set of equations. Considering the spherical Fermi surface, the effective mass, $m^*$, and the mean free path, $l_e$, are estimated by solving the following equations simultaneously as performed in \cite{elec},
\begin{align}
\label{eqn7:gn}
\begin{split}
\gamma_{n} & = \left(\frac{\pi}{3}\right)^{2/3}\frac{k_{\mathrm{B}}^{2}m^{*}V_{\mathrm{f.u.}}n^{1/3}}{\hbar^{2}N_{A}},\\
l_{e} & = \frac{3\pi^{2}{\hbar}^{3}}{e^{2}\rho_{0}m^{*2}v_{\mathrm{F}}^{2}},
\\
n & = \frac{1}{3\pi^{2}}\left(\frac{m^{*}v_{\mathrm{F}}}{\hbar}\right)^{3}
\end{split}
\end{align}
where $k_{\mathrm{B}}$ is the Boltzmann constant, $V_{\mathrm{f.u.}}$ is the volume of a formula unit, and $N_A$ is the Avogadro number. Further, in  the superconducting dirty limit, the effective magnetic penetration depth, $\lambda_{GL}$(0) in terms of London penetration depth $\lambda_L$, reads as,
\begin{equation}
\lambda_{GL}(0) = \lambda_L\left(1+\frac{\xi_{0}}{l_e}\right)^{1/2},
\label{eqn8:lam}\\
\lambda_{L}=\left(\frac{m^{*}}{\mu_{0}n e^{2}}\right)^{1/2}
\end{equation}
where $\xi_{0}$ is the BCS coherence length. For $T$ = 0 K, GL coherence length $\xi_{GL}(0)$ and BCS coherence length $\xi_{0}$ are related by expression,
\begin{equation}
\frac{\xi_{GL}(0)}{\xi_{0}} = \frac{\pi}{2\sqrt{3}}\left(1+\frac{\xi_{0}}{l_e}\right)^{-1/2}.
\label{eqn9:xil}
\end{equation}
\begin{table}[b]
\caption{Parameters in the superconducting and normal state of Nb$_{1.7}$Ta$_{3.3}$S$_2$}
\label{ap}
\begingroup
\setlength{\tabcolsep}{12pt}
\begin{tabular}{c c c} 
\hline\hline
Parameters & Unit & Nb$_{1.7}$Ta$_{3.3}$S$_2$ \\ [1ex]
\hline
$T_{c}$& K& 3.64(2)\\             
$H_{c1}(0)$& mT& 10.53(8)\\               
$H_{c2}^{res}(0)$& T& 1.36(1)\\
$H_{c2}^{P}(0)$& T& 6.77(1)\\
$H_{c2}^{Orb}(0)$& T& 0.63(3)\\
$\xi_{GL}$& nm & 19.2(2)\\
$\lambda_{GL}$& nm & 195(3)\\
$k_{GL}$& & 10.1(2)\\
$\gamma_{n}$&  mJ mol$^{-1}$ K$^{-2}$& 21.7(2)\\
$\theta_{D}$& K& 263(1)\\
$\Delta(0)/k_{\mathrm{B}}T_{c}$&  &  1.70(5)\\ 
$v_{\mathrm{F}}$& 10$^5$m s$^{-1}$& 3.8(3) \\
$n$& 10$^{28}$m$^{-3}$& 2.2(8)\\
$\xi_0$/$l_e$ & & 15.6(2)\\
$T_{\mathrm{F}}$& K& 18305(350)\\
$T_{c}/T_{\mathrm{F}}$& & 0.00002(1)\\
$m^{*}$/$m_{e}$&  & 1.8(7)\\
[1ex]
\hline\hline
\end{tabular}
\endgroup
\end{table}

\begin{figure}
\includegraphics[width=0.95\columnwidth]{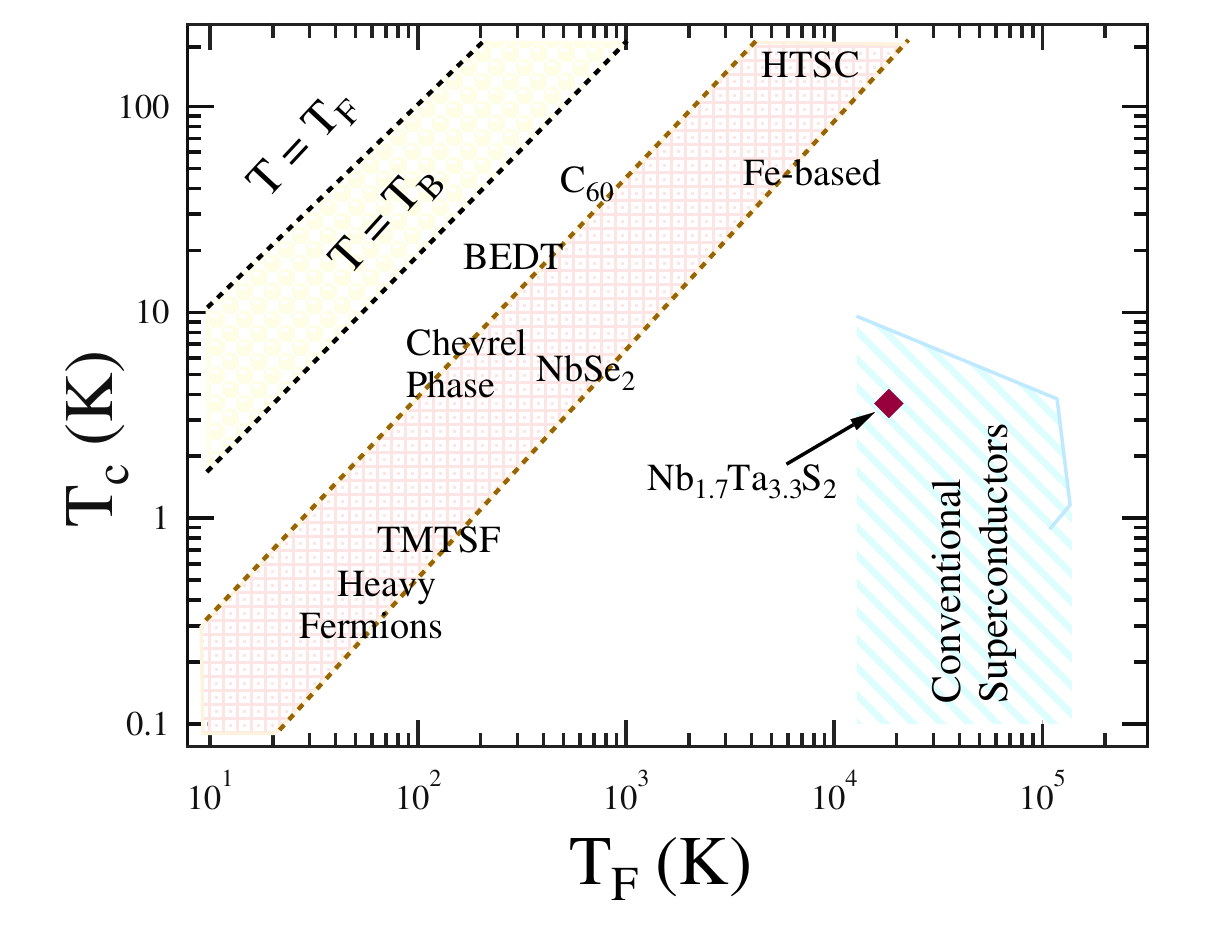}
\caption{\label{Fig4:UP} A plot between the superconducting transition temperature, $T_{c}$ and the effective Fermi temperature, $T_{\mathrm{F}}$. Nb$_{1.7}$Ta$_{3.3}$S$_2$ represented by a solid square marker is shown with other unconventional superconductors families \cite{umera,umera1,umera2,umera3}.} 
\end{figure}
The obtained values of the electronic parameters are listed in \tableref{ap}, and the large ratio $\xi_0$/$l_e$ suggests the dirty limit superconductivity in Nb$_{1.7}$Ta$_{3.3}$S$_2$. In addition, the estimated parameters are used to obtain the Fermi temperature of the system by the expression \cite{Tf},
\begin{equation}
 k_{B}T_{\mathrm{F}} = \frac{\hbar^{2}}{2}(3\pi^{2})^{2/3}\frac{n^{2/3}}{m^{*}}, 
\label{eqn10:tf}
\end{equation}
where $n$ is the quasi-particle number density per unit volume, and $m^{*}$ is the effective mass of quasi-particles. The ratio of $T_{\mathrm{F}}$ and $T_c$ is used to classify a superconductor according to the Uemura scheme and differentiate between the unconventional and conventional superconductors \cite{umera}. The calculated Fermi temperature, $T_{\mathrm{F}}$ = 18305(350) K places Nb$_{1.7}$Ta$_{3.3}$S$_2$ inside the conventional superconductor region as shown in \figref{Fig4:UP}.

\section{CONCLUSION}

The van der Waals metal-rich layered subsulfide Nb$_{1.7}$Ta$_{3.3}$S$_2$ is synthesized, and detailed superconducting properties measurements are performed using AC transport, magnetization and specific heat measurements. It crystallizes in a tetragonal structure with space group $I4/mmm$. Our results confirm a type-II superconductivity with a transition temperature of 3.64(1) K. The sample superconducting and thermodynamic parameters are summarized in \tableref{ap}. Furthermore, the specific heat jump value, $\Delta(0)/k_{\mathrm{B}}T_c$ and the electron-phonon coupling constant, $\lambda_{e-ph}$ are close to the prediction of BCS theory in the weak coupling limit, suggesting the weakly coupled superconducting state in Nb$_{1.7}$Ta$_{3.3}$S$_2$. However, the presence of strong metal interactions, high SOC strength, and a van der Waals layered structure does not appear to impact superconducting behaviour significantly. As these subsulfide layered materials are potential candidates for the superconducting diode effect, it is crucial to fully understand the role of the layered structure, the number of metal layers and the chalcogen stoichiometry in metal-rich van der Waals layered superconducting compounds, further experiments using single-crystalline samples and low-temperature and microscopic measurements are needed.

\section{Acknowledgments}
A. Kataria acknowledges the funding agency Council of Scientific and Industrial Research (CSIR), Government of India, for providing SRF fellowship (Award No: 09/1020(0172)/2019-EMR-I). R. P. S. acknowledge the Science and Engineering Research Board, Government of India, for the Core Research Grant CRG/2019/001028.

\section{Appendix}

\begin{figure}
\includegraphics[width=1.0\columnwidth, origin=b]{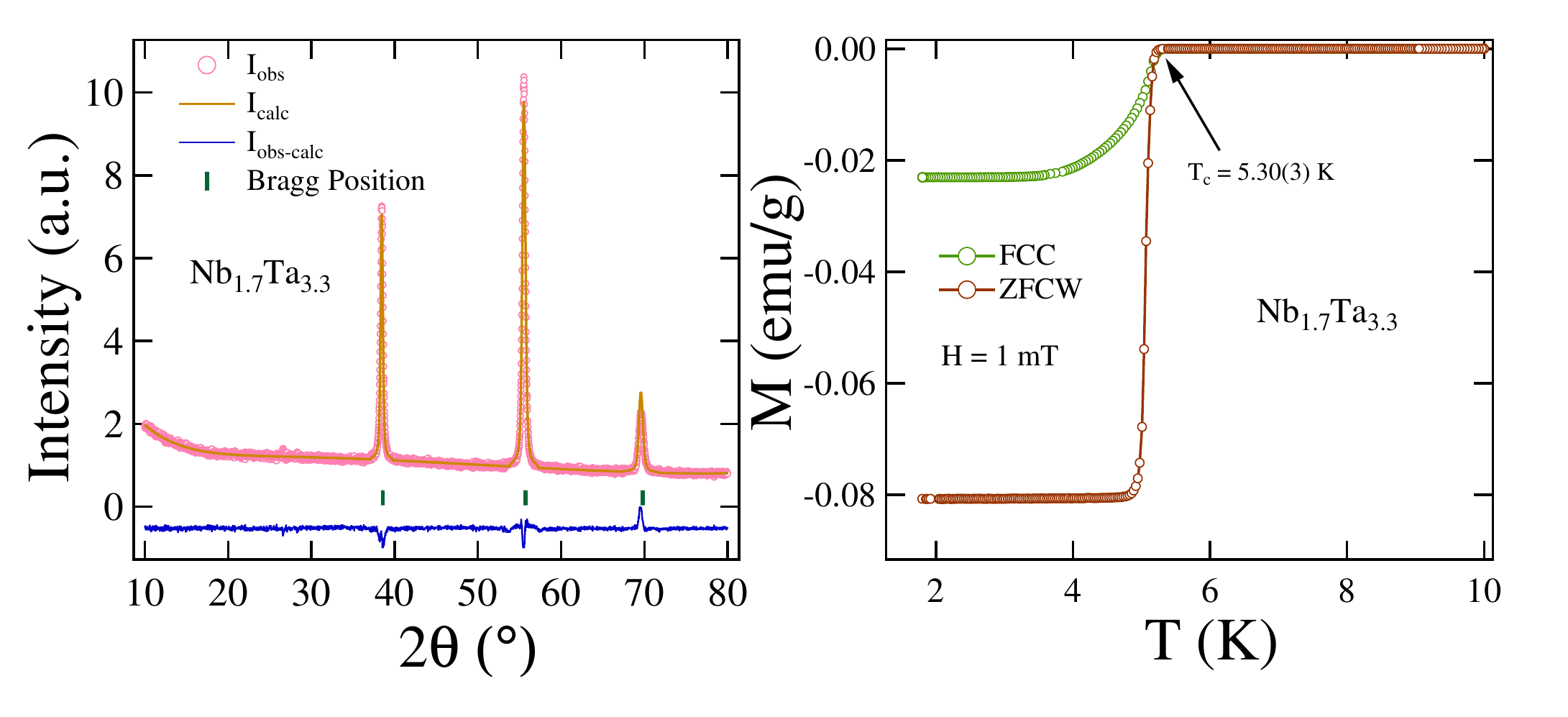}
\caption{\label{Fig5:nbta} (a) Refined XRD pattern of the Nb$_{1.7}$Ta$_{3.3}$ alloy where solid brown line represents the calculated pattern and vertical green bar are the Bragg positions. (b) The temperature variation of magnetic measurement under 1 mT of the alloy. }
\end{figure}

The metallic layer Nb$_{1.7}$Ta$_{3.3}$ also demonstrates superconducting properties. To gain further insight into the superconducting characteristics of Nb$_{1.7}$Ta$_{3.3}$S$_2$ from its metallic section, we produced Nb$_{1.7}$Ta$_{3.3}$ alloy using the standard arc melting method. The powder XRD pattern of the as-cast sample was collected at room temperature. The refined XRD pattern verified that it crystallizes in the space group $Im$-$3m$ BCC structure, with lattice parameter $a$ = $b$ = $c$ = 3.3084(3) \text{\AA}, as depicted in \figref{Fig5:nbta}(a). Furthermore, we measured the alloy using the two protocols, ZFCW and FCC, in a 1 mT magnetic field. In \figref{Fig5:nbta}(b), we identified a distinct diamagnetic onset at $T_c$ = 5.30(3) K, which is indicative of superconductivity. The significant contrast in crystal structure and transition temperature negates the possibility of Nb$_{1.7}$Ta$_{3.3}$ alloy demonstrating superconductivity in Nb$_{1.7}$Ta$_{3.3}$S$_2$.

\end{document}